\documentclass{iaus}
\usepackage{graphicx}

\title[JD 11.~~Coronal Radio Remote Sensing] 
{Radio Remote Sensing of the Corona\\ and the Solar Wind}

\author[Steven R. Spangler \& Catherine A. Whiting]   
{Steven R. Spangler
\and Catherine A. Whiting}

\affiliation{Department of Physics and Astronomy, University of Iowa, Iowa City, Iowa, 52242, USA\\ email: {\tt steven-spangler@uiowa.edu} }

\pubyear{2008}
\volume{257}  
\pagerange{1--11}
\jname{Universal Heliophysical Processes}
\editors{N. Gopalswamy \& D. Webb, eds.}
\begin{document}

\maketitle

\begin{abstract}
Modern radio telescopes are extremely sensitive to plasma on the line of sight from a radio source to the antenna.  Plasmas in the corona and solar wind produce measurable changes in the radio wave amplitude and phase, and the phase difference between wave fields of opposite circular polarization.  Such measurements can be made of radio waves from spacecraft transmitters and extragalactic radio sources, using radio telescopes and spacecraft tracking antennas.  Data have been taken at frequencies from about 80 MHz to 8000 MHz.  Lower frequencies probe plasma at greater heliocentric distances.  Analysis of these data yields information on the plasma density, density fluctuations, and plasma flow speeds in the corona and solar wind, and on the magnetic field in the solar corona. This paper will concentrate on the information that can be obtained from measurements of Faraday rotation through the corona and inner solar wind.  The magnitude of Faraday rotation is proportional to the line of sight integral of the plasma density and the line-of-sight component of the magnetic field.  Faraday rotation provides an almost unique means of estimating the magnetic field in this part of space.  This technique has contributed to measurement of the large scale coronal magnetic field, the properties of electromagnetic turbulence in the corona, possible detection of electrical currents in the corona, and probing of the internal structure of coronal mass ejections (CMEs).  This paper concentrates on the search for small-scale coronal turbulence and remote sensing of the structure of CMEs.  Future investigations with the Expanded Very Large Array (EVLA) or Murchison Widefield Array (MWA) could provide unique observational input on the astrophysics of CMEs.  
\end{abstract}

\section{Introduction}
Understanding the physics of the solar corona and its transition to the solar wind requires specifying the plasma physics properties in the region from the coronal base to heliocentric distances of tens of solar radii. We require knowledge of parameters such as the plasma density, vector magnetic field, plasma flow velocity, and ion and electron temperatures. Properties of turbulence are also important, so that we may test various theories which invoke turbulence as an important agent.  

It is in this part of space that radioastronomical propagation measurements can play an important role.  The basic idea is to observe a radio source which is viewed through the corona.  This radio source can be an extragalactic radio source like a radio galaxy or a quasar, or the transmitter of a spacecraft.  A cartoon illustrating the geometry is shown in Figure 1 of \cite{Spangler02}. The radio waves propagate through the corona or the inner solar wind, and are modified in their amplitude and phase.  A radio telescope on Earth can measure these amplitude or phase changes, and infer properties of the intervening plasma. Further details are discussed in the literature; illustrative papers are those of \cite{Bird90} and \cite{Spangler02}.

This paper will emphasize observation of Faraday rotation in the plasma of the outer corona and inner solar wind.  A linearly polarized radio wave propagating through a magnetized plasma will undergo a rotation in the plane of linear polarization, with the change in polarization position angle $\Delta \chi$ given by
  \begin{equation}
\Delta \chi = \left[ \left( \frac{e^3}{2 \pi m_e^2 c^4}\right) 
              \int_L n_e \vec{B} \cdot \vec{dz} \right] \lambda^2
\end{equation}
Equation (1) is in cgs units.  The term in square brackets on the right hand side is referred to as the {\em rotation measure} ($RM$).   It is conventionally reported in SI units of radians/m$^2$; the cgs value of the rotation measure is converted to SI by multiplying by a factor of $10^4$.

The geometry of coronal Faraday rotation is illustrated in Figure 1 of \cite{Spangler05}.  Radio waves from a linearly polarized radio source, in this case a radio galaxy or quasar, propagate towards the Earth and pass close to the Sun.  Because of the higher plasma density and magnetic field strength close to the Sun, the measured $RM$ is mainly determined by plasma close to the ``proximate point'', i.e. the point on the line of sight which is closest to the Sun.  The heliocentric distance of the proximate point is called the ``impact parameter''.  

Our investigations (e.g. \cite{Sakurai94,Mancuso99,Mancuso00,Spangler00,Spangler05,Ingleby07,Spangler07}) have used the Very Large Array (VLA) of the National Radio Astronomy Observatory to make such measurements.  With the VLA, there are many radio galaxies and quasars which are sufficiently strong to make these measurements. An observation of one or more such sources juxtaposed to the Sun can be made almost every day.  Radio galaxies and quasars radiate by synchrotron radiation, and are linearly polarized at the level of several percent or more. 

\section{The Magnitude of Coronal Faraday Rotation}
Independent estimates of the plasma density in the inner heliosphere exist and allow us to make rough estimates of the Faraday rotation as a function of impact parameter and frequency of observation.  These estimates obviously require knowledge of the strength and form of the coronal magnetic field as well.  For the most part, these magnetic field estimates result from prior Faraday rotation measurements.   

We assume that the plasma density (strictly speaking, electron density) is a function only of the heliocentric distance $r$,
\begin{equation}
n(r) = N_0 \left( \frac{r}{R_{\odot}}\right)^{-2.5}
\end{equation} 
We also assume that the magnitude of the magnetic field is similarly a function only of heliocentric distance, and is radial
\begin{equation}
\vec{B}(r) = B_0 \left( \frac{r}{R_{\odot}}\right)^{-2} \hat{e}_r
\end{equation}
where $ \hat{e}_r$ is the unit vector in the radial direction. The direction of the magnetic field will change with location in the corona.  We introduce a dimensionless function of heliographic coordinates, $m$, to describe the polarity of the coronal field.  In the simplest case, $m=\pm 1$, depending on location in the corona, but in the more general case  
$-1 \leq m \leq 1$.  

We substitute equations (2.1) and (2.2) into (1.1).  As pointed out by \cite{Patzold87}, the resulting integral is simplified if we make a change of variable from $z$, the spatial coordinate along the line of sight, to $\beta$, an angle which is defined by the point along the line of sight, the center of the Sun, and the proximate point (see definition of $\beta$ in Figure 1 of \cite{Ingleby07}). The resultant expression for the Faraday rotation is
\begin{equation}
\Delta \chi = 2.63 \times 10^{-17} R_{\odot} \left[ N_0 B_0 \right] \frac{\lambda^2}{R_0^{3.5}} \int_{-\pi/2}^{\pi/2} d \beta \cos^{5/2}\beta \sin \beta m(\beta) 
\end{equation} 
In equation (4), $R_{\odot}$ is the radius of the Sun, $2.63 \times 10^{-17}$ is the cgs value of the set of fundamental constants in parentheses in equation (1.1), and $R_0$ is the (dimensionless) value of the impact parameter in units of the solar radius. The wavelength of observation is $\lambda$.  The modulation function $m(\beta)$ is now expressed as a function of the angle $\beta$.  

The integral in equation (2.3) is very important in determining the magnitude of the coronal Faraday rotation. As pointed out in \cite{Ingleby07}, if the magnetic field is of constant polarity along the line of sight, and only a function of $r$, $\Delta \chi$ is exactly zero.  This is obvious from the form of the integral.  The maximum value of the integral occurs if 
\begin{eqnarray}
m(\beta) = -1, \beta < 0  \\
m(\beta) = +1, \beta \geq 0 \nonumber
\end{eqnarray}
In this case, the integral has a value of 4/7.  In what follows, and to parameterize nonoptimum coronal conditions, we express the integral as $\frac{4}{7} \epsilon$.  

We adopt representative values of the constants $N_0 = 1.83 \times 10^6$ cm$^{-3}$ and $B_0=1.01$ G (\cite{Ingleby07}).Our observations with the VLA have been made at wavelengths of about 20 cm, and at impact parameters of $5 R_{\odot}$ or more.  To obtain a useful empirical formula, we then let $\lambda = 20 \lambda_{20}$, and $R_0 = 5 R_5$.  

With all of these normalizations and substitutions, we have the following handy, empirical formula for the magnitude of coronal Faraday rotation.
\begin{equation}
\Delta \chi = 158^{\circ} \left[ \frac{\lambda_{20}^2}{R_5^{3.5}}\right] \epsilon
\end{equation} 
It is again worth emphasizing that $|\epsilon | \leq 1$, and in most cases it will be much less than unity.  However, the observations of \cite{Spangler05} showed coronal Faraday rotation about equal to the prediction of equation (2.5) with $\epsilon =1$.  

For some applications, it is more useful to express the rotation measure rather than the position angle rotation.  For the same set of parameters as above, the coronal rotation measure is  
\begin{equation}
RM = \frac{69 \epsilon}{R_5^{3.5}} \mbox{ rad/m}^2
\end{equation}
Equations (2.5) and (2.6) will be used in Section 8 when we discuss Faraday rotation measurements at greater heliocentric distances. 

An appealing feature of coronal Faraday rotation (in contrast to the case for Faraday rotation in interstellar and ionospheric plasmas) is that the reference observations in the absence of the coronal plasma are easily made.  Observations of the source are made when the line of sight passes through the corona, and then later when the Sun has moved away from that part of the sky.  With relatively rare exceptions, extragalactic radio sources do not change their polarization properties on timescales of a few weeks.  The maps of polarization position angle can then be differenced to yield the coronal Faraday rotation.  The process is illustrated in Figure 3 of \cite{Ingleby07}. 
\section{Information Obtainable from Coronal Faraday Rotation} 
Observations of coronal Faraday rotation provide valuable, and often unique constraints on properties of the coronal plasma.  A partial list of coronal properties which may be inferred is as follows. 
\begin{itemize}
\item {\bf The Large Scale Coronal Magnetic Field.} Measurements on a large number of lines of sight over a range in the impact parameter $R_0$ produce our best model for the coronal magnetic field at heliocentric distances of $ \sim 5 - 10 R_{\odot}$.  Results in this area are given in \cite{Patzold87}, \cite{Mancuso00}, and \cite{Ingleby07}.  
\item {\bf MHD Turbulence in the Corona.} Like all astrophysical fluids, the corona should be turbulent.  This turbulence may play an important role in the thermodynamics of the corona.  The stochastic fluctuations in plasma density and magnetic field which occur in plasma turbulence generate corresponding fluctuations in the coronal Faraday rotation.  Measurements of, or limits to these fluctuations have been reported by \cite{Hollweg82}, \cite{Sakurai94}, \cite{Efimov93}, \cite{Mancuso99}, and \cite{Spangler00}, among others. Although Faraday rotation fluctuations have been measured which are probably due to magnetohydrodynamic turbulence, it not certain at this point whether this turbulence has a sufficient energy density and dissipation rate to make a significant contribution to coronal heating. 
\item {\bf Detection of Electrical Currents in the Corona.} Electrical currents must flow in the solar corona to produce the intricate structure seen there.  They may also play a role in coronal heating via Joule Heating. Nonetheless, remote measurement of these currents is very difficult.  \cite{Spangler07} pointed out that VLA observations of an extended radio source through the corona (so that simultanous lines of sight on two or more paths can be measured) can yield information on these currents.  The observational signature of currents is {\em differential Faraday rotation}, or a difference in the rotation measure on two closely juxtaposed lines of sight.  \cite{Spangler07} reported just such observations during VLA measurements of the source 3C227 in August, 2003. The inferred electric current between the two lines of sight was $2.5 \times 10^9$ Amperes in the clearest case, with smaller values for, or upper limits to the current in the remainder of two days of observation. 
\item {\bf Internal Structure of Coronal Mass Ejections.}  If a coronal mass ejection (CME) crosses the line of sight to a polarized radio source, the resulting Faraday rotation can be used to extract information on the magnetic field and plasma density in the interior of the CME.  This can provide unique information about the structure of these objects close to the Sun and in interplanetary space.  Further discussion of this topic is given in Section 6 below.   
\end{itemize}
\section{The Spectrum of Turbulence in the Solar Corona}
Central to the question of turbulent heating of the corona is the spatial power spectrum of the turbulence. It is usually assumed that this spectrum is a power law extending from the outer scale of several tenths of a solar radius to a few solar radii, to an inner scale comparable to plasma microscales such as the ion inertial length.  Depending on location in the corona, the inner scale should be a few kilometers.  

This assumption about the coronal spectrum is based on observed astrophysical power spectra, such as the turbulence spectrum in the solar wind at the location of the Earth, or the inferred spectrum for the interstellar medium. They possess this power law nature, with an inertial range of a few to many decades.  A good illustration of the power law spectrum of magnetic field fluctuations in the solar wind is given in Figure 1 of \cite{Bavassano82}. 

However, it is not certain that the turbulence power spectrum in the corona must be of this power law form.  There could be an enhancement of power closer to the plasma microscales, generated by plasma kinetic processes, and occurring in a part of wavenumber space where dissipation processes are more active. Such turbulence would have a larger energy density on small scales where the dissipation rate is higher, and thus have a larger volumetric heating rate.  

A well-known example of turbulence with these properties is the wave environment upstream of the quasi-parallel portion of the Earth's bow shock. Power spectra of the magnetic field components showing substantial enhancements on small spatial scales over the solar wind background are shown, for example, in Figure 4 of \cite{Spangler97}.  These quasi-monochromatic fluctuations correspond to obliquely-propagating fast mode magnetosonic waves generated by an ion streaming instability.  
There have been theoretical suggestions that enhanced turbulence on short spatial scales also exists in the corona, the dissipation of which could be responsible for heating of the corona (\cite{McKenzie95}, \cite{Marsch97})        

\section{Faraday Screen Depolarization: A Diagnostic for Small Scale Coronal Turbulence}
A radioastronomical effect called {\em Faraday Screen Depolarization} can detect the presence of small-scale, intense turbulence in the solar corona (\cite{Spangler00}).  The physical concept is illustrated in cartoon form in Figure 1 of that paper. Briefly stated, if intense, small-scale turbulence is present in a plasma medium between an imaging radio telescope and an extended, polarized radio source, the turbulence will produce small scale randomization of the polarization position angle.  Averaging over the beam of the radio telescope will then cause a drop in the degree of linear polarization which is measured.  This phenomenon is probably responsible for the well-established observational fact that radio sources are less strongly linearly polarized at low radio frequencies than high.  

\cite{Spangler00} presented formulas relating the drop in the degree of linear polarization to properties of the turbulent screen in the corona.  The analysis identified two parameters which governed the depolarization, or decrease in the degree of polarization.  The first is $x \equiv l_0/L_f$, where $l_0$ is the outer scale of the turbulence, and $L_f$ is the ``footprint'' of the radiotelescope beam.  The radiotelescope beam subtends an angle which corresponds to a physical size in the corona; this is $L_f$.  The second parameter is the asymptotic value of the rotation measure structure function, $D_{\infty}$.  This is the value of the rotation measure structure function which would be measured for spatial (or angular) lags greater than the outer scale.  It is twice the variance in the rotation measure introduced by the turbulent screen.  

The depolarization $D$ is defined and given by (\cite{Spangler00}, equation (8))
\begin{equation}
D^2 \equiv \frac{m^2}{m_0^2} =  \frac{K}{2} \left[ \frac{x^2}{\xi} \left( 1 - e^{-\xi} \right) + \frac{2}{K}e^{-\xi}  \right]
\end{equation}
In equation (5.1) $m$ is the observed degree of linear polarization of a radio source (or part of a radio source) when viewed through the turbulent screen, and $m_0$ is the corresponding intrinsic degree of linear polarization, i.e. that which would be measured in the absence of the screen.  The variable $\xi$ is defined as
\begin{equation}
\xi \equiv \frac{Kx^2}{2} + 2 \lambda^2 D_{\infty}
\end{equation}
where $K \equiv 4 \log 2$ and $\lambda$ is the wavelength of observation.  

\cite{Spangler00} analysed existing observations for evidence of screen depolarization, and found none.  The above formulas were used to place constraints on the properties of small scale turbulence in the corona.  The results of those calculations are given in Sections 4 and 5 of \cite{Spangler00}.  

Subsequently, better observations were made with this type of analysis in mind.  The radio source 3C228 was observed on August 16 and August 18, 2003.  Results from these observations have been presented in \cite{Spangler05}, which shows a radio image of the source and describes its polarization characteristics (Figure 2 of \cite{Spangler05}).  3C228 is a radio galaxy with a double structure, consisting of two highly polarized ``hot spots'' separated by about 46 arcseconds, corresponding to a linear separation of 33,000 km in the corona. The footprint $L_f$ for these observations was determined by a convolution of the interferometer beam and the angular broadening function due to coronal scattering (itself a consequence of small-scale coronal turbulence).  The net footprint was about 4000 km. Simultaneous observations were made at frequencies of 1465 and 1665 MHz.  The source was observed for a period of 8 hours, with the basic observational unit consisting of a scan of 15 minutes duration.  A depolarization measurement was made for each scan at both frequencies of observation.

The results of our observations on August 16, 2003 are shown in Figure 1. The different symbols indicate measurements of the two source components at the two frequencies of observation.  
\begin{figure}[h]
\begin{center}
\includegraphics[angle=-90,width=3.4in]{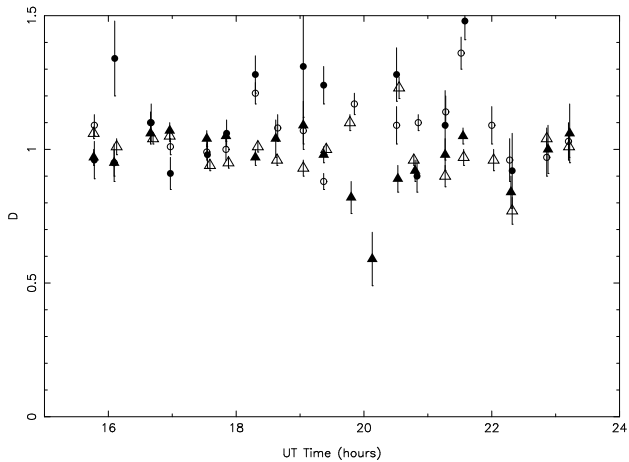} 
 \caption{Measurements of depolarization of the source 3C228 on August 16, 2003.  Each plotted data point represents a measurement of the depolarization $D$ (see equation (5.1) for definition) resulting from a single, 15 minute scan.  Circles and triangles distinguish measurements for the two hot spots in the source.  Filled symbols represent measurements at 1465 MHz, open circles are those at 1665 MHz.}
   \label{fig1}
\end{center}
\end{figure}
The data from Figure 1 are consistent with $D=1$ throughout the observation, i.e. there is no evidence for screen depolarization.  There is a single low point at 20 UT.  The scan from which this measurement was made was of marginal quality, and the depolarization is not confirmed by adjacent scans at 18cm.  The large scatter in the measurements results from the noise-sensitivity of the depolarization measurement; both $m$ and $m_0$ are obtained from a ratio of two measurements (that of the polarized intensity and the total intensity), each of which has its characteristic measurement errors.  There is no indication of a {\em systematic}, session-averaged offset of $D$ from 1.0.  During the observing session, the rotation measure to the source increased (\cite{Spangler05}) and the line of sight passed deeper into a coronal streamer.  It is therefore significant that there is no trend towards smaller $D$ as the session progressed.  Similar results (i.e no evidence for depolarization) were obtained on August 18, 2003, when the impact parameter of the observations was even smaller. 

An analysis similar to that in \cite{Spangler00} has been made on these new data, and new limits obtained for the properties of small scale turbulence in the corona.  These results are shown in Figure 2. The model curves were calculated in a way similar to those in \cite{Spangler00}, and employ models for the background coronal plasma properties similar to those written in equations 2.1 and 2.2.

\begin{figure}[b]
\begin{center}
\includegraphics[angle=-90,width=3.4in]{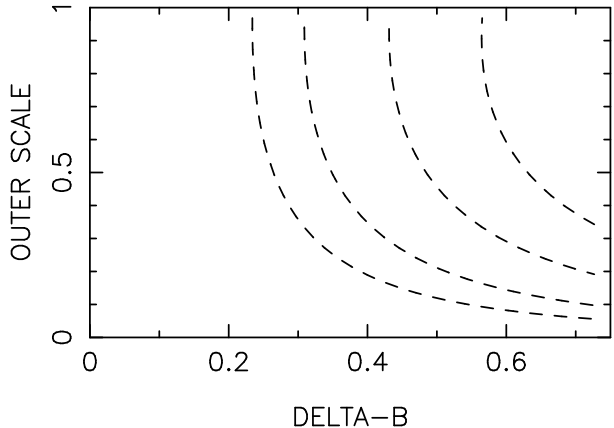}  
 \caption{Range of parameter space allowed by the depolarization measurements of 3C228 on August 16 and 18, 2003. The abscissa of the parameter space is $\frac{\delta b}{B_0}$, where $\delta b$ is the rms value of the fluctuating magnetic field in the turbulence, and $B_0$ is the mean magnetic field. The ordinate is $\frac{l_0}{L_f}$ (ratio of outer scale of turbulence to beam footprint).  The dashed curves correspond to different values of the reduced $\chi^2_{\nu}$.  The curves correspond to reduced chi-square values of 1.90, 2.30, 3.30, and 5.00, progressing from left to right. The first two curves correspond to chi-squared probability of occurrence of 5 \% and 2 \%, respectively. The part of parameter space below and to the left of a curve represents parameters which are acceptable at that level.   }
   \label{fig2}
\end{center}
\end{figure}

Figure 2 shows that a wide range of possible models for small scale turbulence are compatible with our depolarization measurements.  However, there are constraints which can be imposed. Turbulence with an outer scale of order the beam footprint $l_0 = 0.5 - 1.0 L_f$ is compatible with the observations, but the turbulent amplitude $\frac{\delta b}{B_0}$ in that case must be of order $\frac{\delta b}{B_0} \leq 0.2 - 0.3$.  Similarly, large turbulent amplitudes could be present in the corona ($\frac{\delta b}{B_0} \geq 0.5$, but the outer scale would be have to be quite small ($l_0 \leq 0.1 - 0.2 L_f$). 

To summarize, our basic observational result is that the corona does not depolarize radio sources, at least at the level probed in the observations to date. A number of mathematical models for small scale coronal turbulence are still compatible (i.e ``not inconsistent'') with the data. However, it is interesting, within the context of a general exploration of turbulence in the universe, that there is no observational indication of coronal turbulence with a spectrum like that seen upstream of the Earth's bow shock.

\section{Faraday Rotation as a Remote Diagnostic of Coronal Mass Ejections} 
Coronal Mass Ejections (CMEs) are among the most interesting objects in heliospheric plasma physics.  A static structure like a solar prominence loses its stability and expands at supersonic speeds into the corona and interplanetary medium.  CMEs are the most important agents in space weather, since their interaction with Earth is the cause of pronounced terrestrial response to solar activity. We would like a better knowledge of the internal plasma properties of CMEs at all points between liftoff and the orbit of the Earth, so that tests may be made of the numerous theoretical descriptions of these objects.  
Faraday rotation is an ideal diagnostic for CMEs, since it provides information on both the density and magnetic field. Some of the questions we would like to ask about CMEs are as follows. 
\begin{itemize}
\item What is the internal plasma structure of a typical CME?
\item Are CMEs approximately describable as force-free flux ropes?
\item What is the physical significance of the ``three part structure'', i.e. an outer loop, a middle void, and an inner loop or core, which is generally observed for these objects?
\end{itemize}

In a recent study, \cite{Liu07} calculated the expected Faraday rotation signatures of various models for CMEs. The simplest such model is that of a force-free flux rope.  The calculated RM for this model is shown in Figure 6 of \cite{Liu07}.  It shows an RM perturbation with an amplitude of about 9 rad/m$^2$ for a structure at a heliocentric distance of 10 $R_{\odot}$, comparable to what is observed. In this model, regions of the sky with opposite sign of the RM are closely juxtaposed.  As discussed by \cite{Liu07}, this is a consequence of the helical magnetic field structure in a force-free flux rope.  \cite{Liu07} note that the helicity of the flux rope and the orientation of the flux rope axis to the line of sight could be determined by RM observations.  

To date, the only high quality set of  observations of RM anomalies due to a CME are those of \cite{Bird85}, which were of occultations of the Helios spacecraft transmitter by CMEs.   \cite{Liu07} describe qualitative agreement between their models and the observations of \cite{Bird85}.

\section{CME Observations with the VLA} The Very Large Array is well-suited for observations of CMEs.  Among the desirable features of the VLA for this type of observation are 
\begin{itemize}
\item the capability of making simultaneous observations at 1465 and 1665 MHz, and thus confirming the $\lambda^2$ dependence of the position angle rotation,
\item relative easy measurement of RMs as small as $\sim 1 \mbox{ rad/m}^2$,
\item imaging of extended radio sources, which allows measurements of differential Faraday rotation through the CME,
\item the option of observing a number of sources on a given day which are available for occultation by a CME.
\end{itemize}

\cite{Ingleby07} briefly describe a CME observation with the VLA, and we now consider those results in more detail.  Some of the observations of \cite{Ingleby07} were made on March 12, 2005.  On that day, there were 7 sources arranged in a particularly good ``constellation'' around the Sun (see Figure 4 of \cite{Ingleby07}).  A coronal mass ejection occurred on the east limb of the Sun, and moved in the direction of two of the sources, 2335-015 and 2337-025.  The outer loop of the CME was in the vicinity of the two sources near the end of the observing session.  

Faraday rotation data for these two sources are shown in Figures 3 and 4. Plotted are the residual polarization position angles as a function of time.  By residual position angles, we mean that the mean $\chi$ for the session has been subtracted from the position angle time series.  This permits us to study variations in the polarization position angle as a function of time.  Measurements at both 1465 MHz (solid symbols) and 1665 MHz (open symbols) are shown. The residuals at 1665 MHz have been multiplied by the square of the ratio of the observing wavelengths, $(\lambda_{1465}/\lambda_{1665})^2$,where $\lambda_{1465}$ is the wavelength corresponding to 1465 MHz, and $\lambda_{1665}$ is that corresponding to 1665 MHz.  Since Faraday rotation is proportional to the square of the wavelength, after this adjustment the measurements at the two frequencies should agree. 

The data for 2335-015 show no credible variation during this period.  There is no significant departure or trend from zero that appears in both the 1465 and 1665 MHz data.  However, the observations of 2337-025 do show a temporal variation in the Faraday rotation.  A progressive change is seen in the last two scans, and the 1465 and 1665 MHz data are in agreement. The total change in position angle (defined as the difference between the position angle measured in the last scan, and the average position angle from 16 - 21h UT) is about $28^{\circ}$, corresponding to a rotation measure of about 12 rad/m$^2$. This is comparable in magnitude to the model value of \cite{Liu07} of 9  rad/m$^2$ at a somewhat larger impact parameter (10 $R_{\odot}$ for the model calculations, compared to  6.6 $R_{\odot}$ for the observations). 
\begin{figure}[h]
\begin{center}
\includegraphics[angle=-90,width=3.4in]{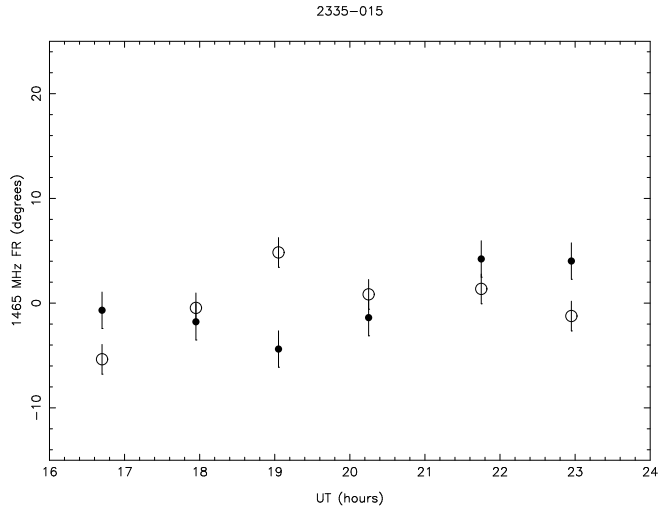}  
 \caption{Faraday rotation observations of radio source 2335-015 on March 12, 2005.  Plotted is the residual polarization position angle (position angle minus the mean for the whole eight-hour session) as a function of UT.  Solid symbols are the measurements at 1465 MHz, open symbols are measurements at 1665 MHz, with the 1665 MHz residual multiplied by the square of the ratio of the observing wavelengths.  No definite evidence is seen of a Faraday rotation event associated with the approach of the CME. }
   \label{fig3}
\end{center}
\end{figure}

\begin{figure}[h]
\begin{center}
\includegraphics[angle=-90,width=3.4in]{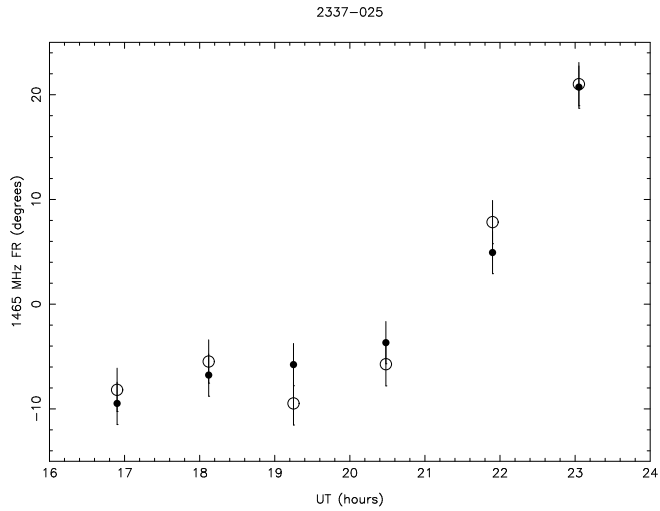} 
 \caption{Faraday rotation observations of radio source 2337-025 on March 12, 2005. Format is the same as Figure 4. For this source, a change in the Faraday rotation is seen in the last 2 scans, which is plausibly attributed to the approaching CME.}
   \label{fig4}
\end{center}
\end{figure}

\begin{figure}[h]
\begin{center}
\includegraphics[width=3.4in]{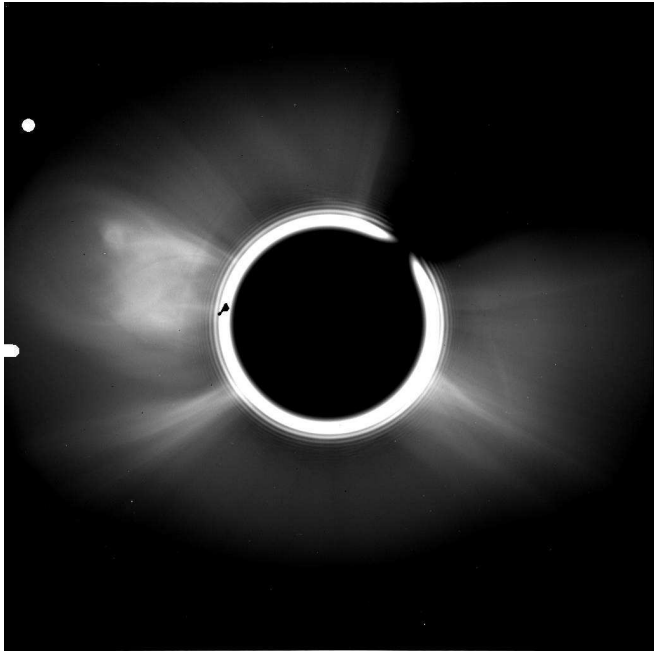} 
 \caption{C2 coronagraph image at 23:12 UT on March 12, 2005.  The white dots 
indicate the positions of the radio sources 2335-015 (upper) and 2337-025 (lower)}
   \label{fig5}
\end{center}
\end{figure}

A curious result of these observations is that it appears the main part of the CME had not occulted the line of sight to the source when the RM variation began.  The data shown in Figure 4 indicate that a plasma structure with detectable RM first occulted the line of sight around 21h, and certainly before 22h.  In the remaining one to two hours of the session, the line of sight appears to have penetrated deeper in the structure.  However, coronagraph observations from the LASCO C2 coronagraph seem to indicate that the ``outer loop'' of the CME crossed the line of sight to the source after the end of the session.  This is illustrated in Figure 5, which shows a LASCO C2 coronagraph image for 23h12m UT on the day of observation.  This image was taken from the LASCO data archive.  The white dots at the left edge of the image indicate the positions of the two radio sources.  The lower of the two indicates the position of 2337-025; it is indicated as an arrow to signify that the position of the source is slightly outside the frame shown.  This image appears to show that at the time of the coronagraph observation, and after the last VLA measurement, the outer loop of the CME had not quite occulted the line of sight.

If the simple astrometry used in generating this figure is correct, we can conclude that there is a magnetohydrodynamic precursor associated with the CME, ahead of the outer loop, and capable of producing an easily measurable variation in the RM.  
\section{Future Faraday Rotation Studies of the Heliosphere} 
The scientific results described here as well as in the papers we have referenced illustrate the uniqueness and utility of Faraday rotation observations for studies of the plasma physics of the solar corona and inner solar wind.  Furthermore, enhanced capability in this field will be available in the very near future due to the availability of the Expanded Very Large Array (EVLA) of the US National Radio Astronomy Observatory and the Murchison Widefield Array, developed by a consortium of the Massachusetts Institute of Technology and several Australian universities.  This enhanced capability will benefit general studies of the heliosphere and its turbulence, and should be of particular interest in new studies of coronal mass ejections.  These future studies offer the potential of an improvement on the study of \cite{Bird85} in two ways.  First, better coronagraph are now available in space in the form of SOHO/LASCO and STEREO/SECCHI.  These coronagraphs provide a better view of the way in which the line of sight to the radio source passes through the CME structure, as well as information on the line-of-sight integral of the electron density.  Second, new radio observations will permit simultaneous observations along multiple lines of sight.  They will also allow differential Faraday rotation measurements which will diagnose the internal structure of the CME.  Finally, future CME measurements may be possible at greater heliocentric distances than in the past.  This would  permit a study of how the CME structure evolves with distance in the interplanetary medium. 

\subsection{The Expanded Very Large Array} The Very Large Array is the premier radio synthesis telescope in the world.  It is presently in the process of a major upgrade in the antennas, receivers, and electronics.  When completed in roughly 2010, it will have greater sensitivity and frequency agility.  This will benefit coronal Faraday rotation measurements.  Since our research program utilizes radio galaxies and quasars for background sources, increased instrumental sensitivity is as important for this type of work as for studies of these objects {\em pro suis}. Improved sensitivity will permit observations of weaker sources than previously, which will increase the density of sources on the sky which can be utilized.  Continuous frequency coverage between 1 and 50 GHz will permit observations over a wide range of heliocentric distances, most importantly deeper in the corona. 

\subsection{The Murchison Widefield Array} The Murchison Widefield Array (MWA) is in the process of being constructed in Western Australia.  It will operate at low radio frequencies, 80 - 300 MHz.  The appealing feature of such frequencies in the present context is that they should permit Faraday rotation measurements at greater heliocentric distances than are possible with the VLA, an instrument which mainly operates at microwave frequencies.  This is illustrated by the discussion in Section 2, particularly equation (2.5). At a wavelength of 300 cm (corresponding to the low end of the MWA frequency range), a given Faraday rotation $\Delta \chi$ can be measured at a heliocentric distance 4.7 times greater than the distance at which this rotation would be measured at 20cm.  Faraday rotation measurements can be made with the VLA at 20 cm at impact parameters out to about 10 $R_{\odot}$; the same amount of rotation might be measurable with MWA out to $ \sim 50 R_{\odot}$.  This would permit a new class of scientific studies.  

The MWA will face two clear challenges in heliospheric Faraday rotation observations.  The first is the empirical fact that extragalactic radio sources become progressively less linearly polarized with decreasing frequency below 1 GHz. This behavior has been known for decades (e.g. \cite{Kronberg70}).  Simply stated, at frequencies of a few hundred megahertz and lower, radio galaxies and quasars are not very effective radiators of polarized radiation. 

The second issue to arise in MWA Faraday rotation observations is that heliospheric physicists will, perforce, become ionospheric physicists as well.  The Earth's ionosphere produces Faraday rotation which varies with direction in the sky, time of day, and date.  For observations at large distances from the Sun, the ionospheric contribution will be much larger than that due to the interplanetary medium.  The problem can be illustrated with equation (2.6).  At an impact parameter of 30 $R_{\odot}$, the expected solar wind RM should be of order $0.13 \epsilon \mbox{ rad/m}^2$. Again, recall that $\epsilon$ will usually be less than 1.  The ionospheric rotation measure is typically of order $0.5 - 1.0 \mbox{ rad/m}^2$ and can be much higher.  As a not-atypical case, \cite{Ingleby07} encountered ionospheric rotation measures of $2.5 - 3.8 \mbox{ rad/m}^2$ on one of the four days of observation in their investigation. Detection of interplanetary RMs of a few tenths of a rad/m$^2$ in the presence of ionospheric contributions which are larger (sometimes much larger), as well as spatially and temporally variable, will take better diagnosing of the ionosphere than has been previously attempted by radio astronomers.  

\section{Summary and Conclusions}
Faraday rotation measurements on radio waves which have propagated through the solar corona and inner solar wind provide a unique plasma diagnostic on that region of space. Measurements to date have provided information on the coronal magnetic field, the amplitude and spatial power spectrum of MHD turbulence in the corona, estimates of electrical currents, and the plasma structure of coronal mass ejections.  Two new instruments which will become available in the next few years, the Extended Very Large Array (EVLA) and the Murchison Widefield Array (MWA), will permit much better measurements, as well as measurements of a different type.  

\noindent
{\bf Acknowledgements.}  We thank the US National Science Foundation, Division of Atmospheric Sciences, for supporting this research via grant ATM03-54782. We also thank the National Radio Astronomy Observatory for allocating significant amounts of VLA observing time to this research program.

\begin{discussion}

\discuss{Cicero}{Utique sensisti naturam omnium rerum, atque omnes
nobis plane docuisti}
\discuss{Ille Ipse}{Gratias tibi ago, Magister; permultum significat, ut 
tu hunc laborem probes.}

\end{discussion}
\end{document}